\documentclass{optica-article}

\journal{opticajournal} 

\articletype{Research Article}

\begin{document}

\title{Image scanning lensless fiber-bundle endomicroscopy}

\author{Gil Weinberg,\authormark{1} Uri Weiss,\authormark{1} and Ori Katz \authormark{1,*}}

\address{\authormark{1}Institute of Applied Physics, The Hebrew University of Jerusalem, 9190401 Jerusalem, Israel\\}

\email{\authormark{*} Orik@mail.huji.ac.il} 

\begin{abstract*} 
    Fiber-based confocal endomicroscopy has shown great promise for minimally-invasive deep-tissue imaging. Despite its advantages, confocal fiber-bundle endoscopy inherently suffers from undersampling due to the spacing between fiber cores, and low collection efficiency when the target is not in proximity to the distal fiber facet. Here, we demonstrate an adaptation of image-scanning microscopy (ISM) to lensless fiber bundle endoscopy, doubling the spatial sampling frequency and significantly improving collection efficiency. Our approach only requires replacing the confocal detector with a camera. It improves the spatial resolution for targets placed at a distance from the fiber tip, and addresses the fundamental challenge of aliasing/pixelization artifacts. 
    
\end{abstract*}


\section{Introduction}

Flexible fiber-based micro-endoscopes are an effective tool for minimally-invasive deep-tissue imaging. They offer micron-scale resolution with miniature probes having diameters of a few hundred microns, making them suitable for \textit{in-vivo} imaging of neurons and cells \cite{flusberg2005fiber, oh2013optical,accanto2022flexible}. One of the popular implementations for miniature micro-endoscopes relies on multi-core fiber bundles (also known as coherent fiber bundles), where each fiber core serves as one imaging pixel. Conventional imaging using fiber bundles is performed either by placing the bare distal fiber tip in contact or close distance to the target, or by adding a gradient-index (GRIN) lens to the distal tip \cite{knittel2001endoscope,gobel2004miniaturized}. 
Both solutions suffer from significant undersampling due to the low fill factor caused by the large spacing between fiber cores, compared to the core size \cite{oh2013optical} and out-of-focus background.

A common approach to improve the resolution and optical cross-sectioning of fiber-bundle endoscopes 
is through confocal imaging \cite{gmitro1993confocal,delaney1994fiber,lane2000fiber,knittel2001endoscope,jabbour2012confocal}. 
In confocal fiber-bundle-based endoscopy the illumination is scanned over each of the fiber cores, and only the light collected from the illumination core is used to form the image. Thus, each single core serves as an effective confocal pinhole, rejecting out-of-focus light and improving the spatial resolution.
Importantly, the implementation of confocal fiber-bundle endoscope does not require the addition of any distal elements since the scan is performed at the proximal fiber end.

Over the years, many techniques have been suggested and demonstrated to improve lensless fluorescence fiber-bundle endomicroscopy. These methods include structured-illumination \cite{santos2009optically,zhang2020high}, retrieving light-field 3D information from the modal excitation within each core \cite{orth2019optical}, optical fluctuation imaging \cite{shekel2020using}, and recently, advanced wavefront-shaping and image reconstruction approaches \cite{thompson2011adaptive, andresen2013toward, kim2014toward, stasio2016calibration, porat2016widefield, weiss2018two}. However, except for the advanced recent approaches where the fiber bundle effectively serves as the imaging pupil \cite{thompson2011adaptive, andresen2013toward, kim2014toward, stasio2016calibration, porat2016widefield, weiss2018two}, all approaches suffer from aliasing/pixelization artifacts due to the undersampling by the fiber cores at the image plane. In addition, lensless confocal fiber-bundle endoscopy suffers from low collection efficiency when the target is not in close proximity to the distal facet.
Unfortunately, the low effective fill factor of coherent fiber bundles, which leads to image undersampling, is an inherent design limitation required to minimize the cross-talk between neighboring cores \cite{oh2013optical}. 

Here, we demonstrate that both the collection efficiency and spatial sampling frequency can be significantly improved for targets placed at a distance from a lensless fiber bundle by adapting the principles of image scanning microscopy (ISM) \cite{muller2010image} (also known as pixel-reassignment microscopy \cite{sheppard1988optik,sheppard2020pixel}) to lensless fiber-bundle endomicroscopy. 

ISM has emerged as a microscopy technique that improves the signal-to-noise and resolution of confocal microscopes. ISM utilizes the same raster-scanned illumination as in confocal microscopy but replaces the small detection pinhole with a larger pixelated detector array, allowing the use of all signal photons without loss of resolution. In ISM, the resolution is maintained or improved compared to confocal microscopy by reassigning the detected light from each detector (pixel) to the midpoint between the detected and illumination positions. The final ISM image is the sum of the reassigned signals. Mathematically, the pixel-reassignment process is equivalent to spatially scaling by half (shrinking) each raw image from the detector array (Fig.~\ref{fig1}), and summing all scaled raw images. 
Recently, ISM has been used to improve collection efficiency and resolution in confocal Raman microscopy\cite{roider2016high}, two-photon microscopy \cite{sun2018improving}, scanning laser ophthalmoscopy \cite{mozaffari2018versatile}, transmission-matrix based imaging in multi-mode fibers (MMF) \cite{singh2023multiview}, and quantum microscopy \cite{tenne2019super,rossman2019rapid}. 

Here, we combine the raster-scanned illumination of conventional confocal endoscopy \cite{gmitro1993confocal} with camera-based detection to implement ISM in a bundle-based lensless microendoscope. Our straightforward approach tackles two main drawbacks of conventional confocal fiber-bundle endoscopy. First, 
we collect fluorescence signals through all fiber-bundle cores rather than a single core. This proves beneficial for targets placed at a distance from the bundle tip, as the collection efficiency is greatly increased. 
Second, ISM doubles the spatial sampling frequency of the fiber bundle due to the spatial scaling of the pixel-reassignment process (Fig.~\ref{fig1}), addressing the fundamental problem of aliasing and pixelization artifacts of targets located at close distances to the fiber. 

\section{Methods}
\subsection{Principle}

\begin{figure}[ht!]
	\centering
	\includegraphics [width=0.96\textwidth]
	{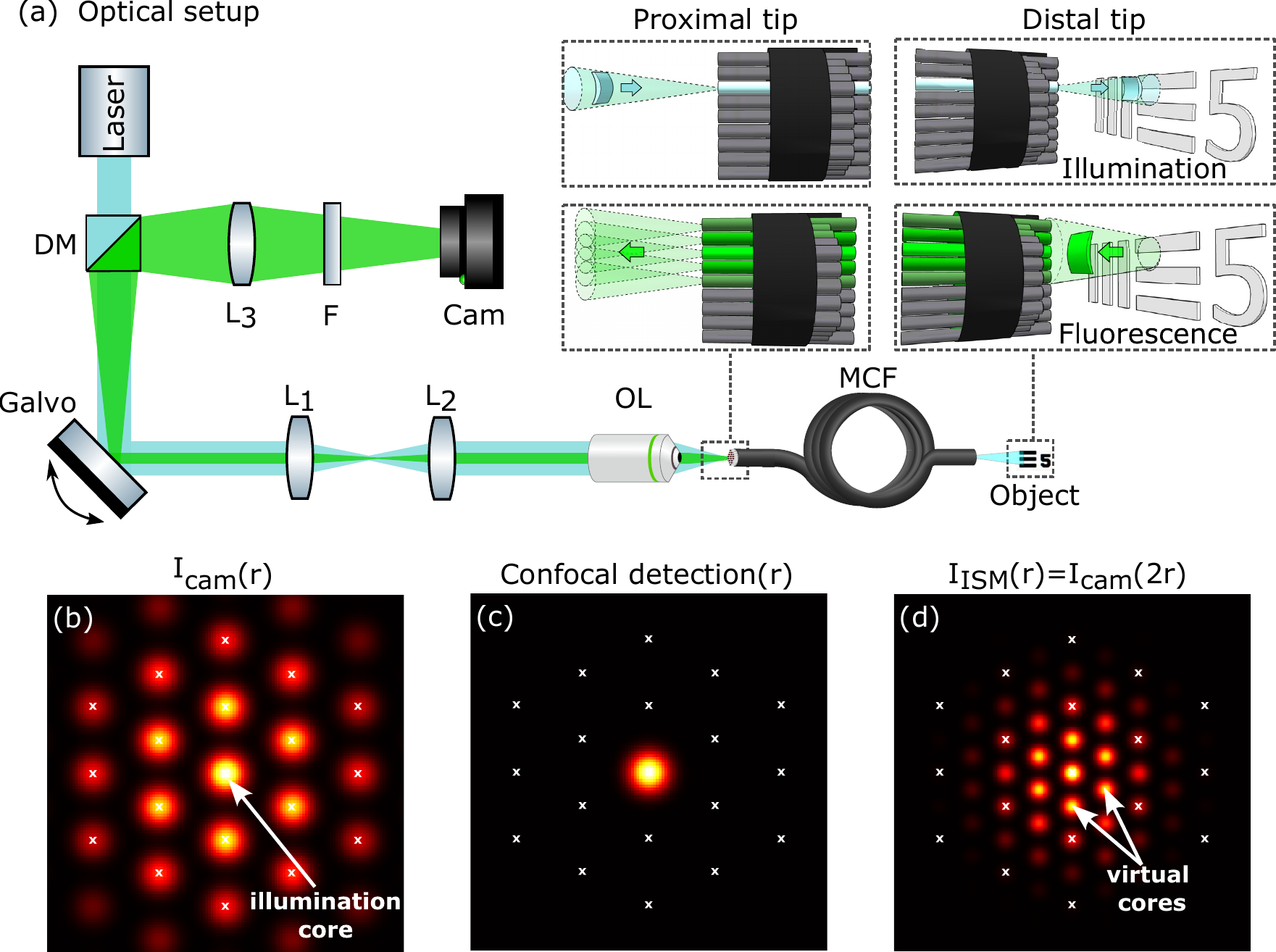}
    	\caption{\textbf{Image scanning lensless fiber-bundle fluorescence endomicroscopy.} (a) setup: At the proximal tip of a multi-core fiber (MCF) a focused laser sequentially illuminates the MCF cores one by one using a galvo-scanner. 
     The excitation light from each core illuminates a target object placed at a small distance from the distal tip (top inset). 
     The fluorescence signal is collected by both the illuminating core and the neighboring cores (bottom inset).
     The collected fluorescence pattern for each illumination is imaged on an sCMOS camera through a dichroic mirror (DM) and a bandpass filter (F). 
     (b) A simulated single camera image for an object placed at a distance of $60 \mathrm{\mu m}$ from the distal tip, where the center core is illuminated. A confocal image can be generated by recording the signal only from the illuminating core, ignoring the signals from the neighboring  cores (c). White 'x's mark the cores positions, i.e. the sampling grid. 
     (d) An ISM image is formed by resizing the full camera image (b) by a factor of $0.5$, and summing all captured frames. Thus, utilizing all detected photons and creating a $\times 2$ denser sampling grid with new 'virtual' cores.}
    	\label{fig1}
\end{figure}

The principle of the image-scanning fiber endoscopy is schematically shown in Fig.~\ref{fig1}. Our approach is based on a conventional confocal endoscopy configuration (Fig.~\ref{fig1}a). A focused beam scans the fiber cores at the proximal tip. The same fiber bundle collects the fluorescence signals. The proximal facet is imaged on a camera through a dichroic beam-splitter (DM) and a band-pass filter (F). At each scan position, the camera captures the fluorescence intensity distribution over the proximal facet (Fig.~\ref{fig1}b). The sampling grid of the bundle cores is marked by white X's in Fig.~\ref{fig1}(b-d). The camera images allow performing both confocal imaging and ISM. For confocal endoscopy only the collected signal from the illumination core is integrated (Fig.~\ref{fig1}c). For ISM, the entire camera image is scaled by a factor of $0.5$ \cite{sheppard2020pixel} and saved  (Fig.~\ref{fig1}d). This reassignment procedure creates 'virtual' cores at a new detection grid, with a spatial sampling frequency of $2 f_s$ that is two times larger than the original cores grid (Fig.~\ref{fig1}b). The final ISM image is the summation of all reassigned images, having a denser spatial sampling grid and a higher collection efficiency. 

Quantitatively, the point spread function (PSF) of confocal imaging, $h_{conf}(r)$, is the 
product of the illumination and the detection PSFs \cite{sheppard2013superresolution}.

\begin{equation}
        h_{conf}(\textbf{r}) =  h_{det}(\textbf{r}) \cdot h_{illum}(\textbf{r})
    \label{eq:conf_psf}
\end{equation}

Where $\textbf{r}$ is the transverse spatial coordinate, $h_{illum}$ is the illumination PSF, and $h_{det}$ is the detection PSF. When both illumination and detection have similar Gaussian profiles, confocal imaging results in a $sqrt(2)$ times narrower effective Gaussian PSF than each of these PSFs.

In the case of ISM, the effective PSF, $h_{ISM}$ is the convolution of the illumination and detection PSFs, scaled by a factor of half \cite{muller2010image}.

\begin{equation}
        h_{ISM}(\textbf{r}) = h_{det}(2\textbf{r}) \ast h_{illum}(2\textbf{r})
    \label{eq:ism_psf}
\end{equation}

In the case of identical Gaussian profiles for illumination and detection PSFs, the final effective PSF of ISM is identical to the confocal. This result assumes an adequately Nyquist sampling of the collected signals. While this is conventionally attained in a conventional microscopy setup, it is not the case in bundle-based endoscopy. 
As a result, the doubling of the spatial sampling frequency in fiber-bundle ISM improves the final optical transfer function (OTF, the Fourier transform of the PSF), even in the case of identical detection and illumination PSFs since it avoids aliasing of spatial frequencies in the range $[f_s/2 f_s]$ (see Fig.~\ref{fig2_sampling}). 

\section{Results}

\subsection{Numerical study of the optical transfer function}

\begin{figure}[ht!]
	\centering
	\includegraphics[width=0.96\textwidth]
	{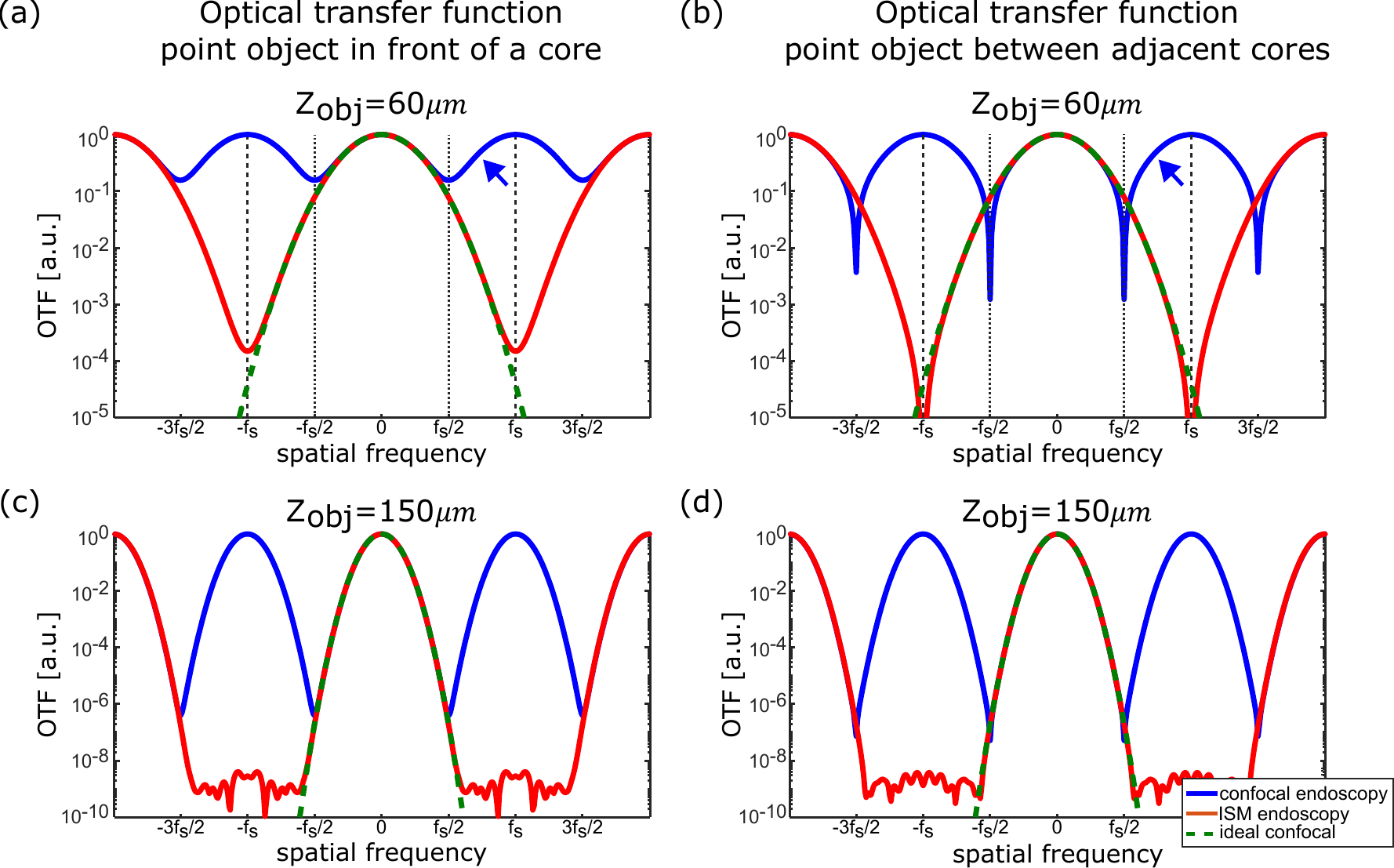}
	\caption{
            \textbf{Simulated OTFs comparisons between confocal endoscopy (blue), ISM endoscopy (red), and confocal imaging without sampling cores (dashed green)} 
            (a,b) OTFs crosssection for an object placed at a distance of  $\mathrm{\mu m}$ from the distal tip: due to the spatial sampling of the MCF cores, endoscopic OTFs suffer from aliasing artifacts exhibited by the folding of high spatial frequencies content (arrows), compared to ideal confocal imaging. 
            ISM-endoscopy (red) doubles the effective spatial sampling frequency from $f_s$ of confocal endoscopy (blue) to $2f_s$ (see Fig.~\ref{fig1}d, and the folding frequencies (dashed and dotted vertical lines, respectively). This allows correct retrieval of $\times 2$ higher spatial frequency content in the ideal unsampled confocal OTF (dashed green). (c-d)  (c-d) Same as (a-b) for an object distance of $150 \mathrm{\mu m}$. At larger object distances, the doubled sampling frequency of ISM endoscopy has a negligible impact due to the lower frequency content of the defocused object image. Nonetheless, the gain in signal collection efficiency of ISM improves the reconstructed images (see Fig.\ref{fig3_experimental} and Supplementary Fig.~S2).}
        \label{fig2_sampling}
\end{figure}

To study the expected improvement in the normalized OTF due to the increase in spatial sampling frequency of ISM fiber-bundle endoscopy, we numerically simulated the resulting images of a point target placed at distances of $60 \mathrm{\mu m}$ and $150 \mathrm{\mu m}$ for ISM and confocal fiber bundle endoscopy, and compared to the ideal confocal image sampled at a spatial frequency $\sim 6$-times that of the fiber bundle (i.e., without relevant aliasing). The OTF for each case was calculated by Fourier transforming the simulated image.
The results of this study are presented in Fig.~\ref{fig2_sampling}. For each distance, we studied the results when the point target is placed directly in front of a fiber core (Fig.~\ref{fig2_sampling}a,c) and when the point target is placed between two neighboring cores (Fig.~\ref{fig2_sampling}b,d). 

As expected, the 1-D cross-sections of the OTFs in all cases show that aliased spatial frequencies, which result from undersampling, are present in the confocal images at frequencies higher than $f_s/2$ (blue curves and arrows). In contrast, in the ISM images, the aliased frequencies are only present at frequencies above $f_s$ (red curves).  
The doubling of the spatial sampling frequency is important at close distances (Fig.~\ref{fig2_sampling}a,b) since the captured image contains higher frequency content than at larger distances (Fig.~\ref{fig2_sampling}c,d), where the effective defocus blur due to the propagation low-pass filters the higher frequencies content. Importantly, as we show below (Fig.~\ref{fig3_experimental}, Supplementary Fig.~S$2$), ISM endoscopy also improves resolution at longer distances due to the significantly improved collection efficiency and SNR.

\subsection{Experimental results}

\begin{figure}[htb!]
	\centering
	\includegraphics[width=0.96\textwidth]
	{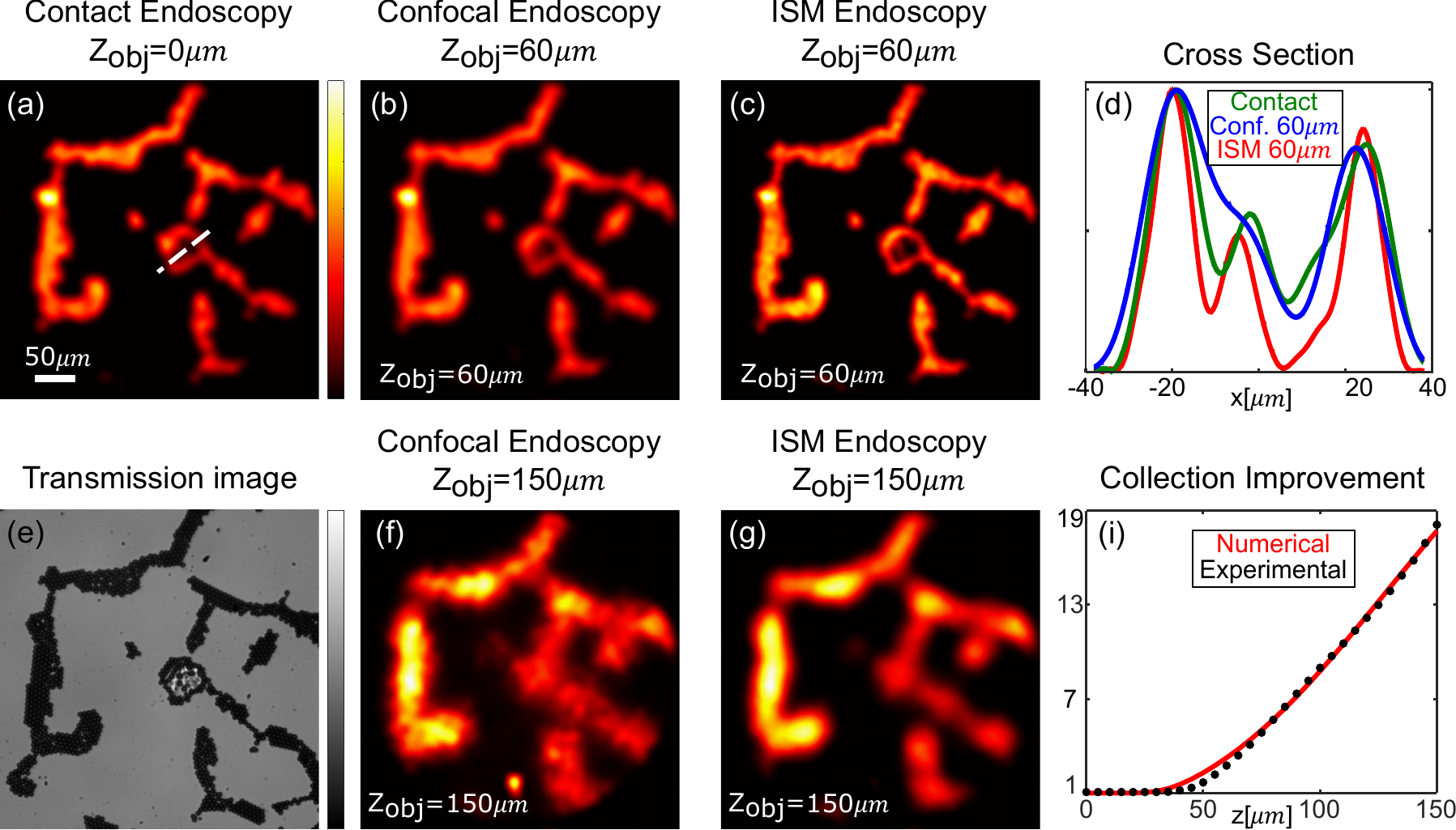}
        \caption{\textbf{Experimental results comparing ISM and confocal endoscopy images of fluorescent beads.} (a) Reference 'contact' image of the fluorescence target when placed in contact with the distal fiber facet ($z_{obj}=0)$. (b) Confocal image of the same target when placed at a distance of $z_{obj}=60 \mathrm{\mu m}$ from the distal tip. (c) ISM image from the same camera frames taken in (b) exhibiting improvements in resolution and SNR. (d) cross-sections of (a-c) over the dashed white line in (a) show the improvement in resolution of ISM endoscopy (red) over confocal endoscopy (blue).  (e) Reference transmission microscope image of the target. (f,g) Same as (b,c) when the target is placed at a distance of $z_{obj}=150 \mathrm{\mu m}$ from the distal tip. At such a large distance, the improvement in SNR of ISM endoscopy due to the higher collection efficiency is significant. 
        (i) Experimental (black dots) and numerical (red trace) plots of the improvement in the signal collection as a function of object distance. The improvement is significant at larger distances due to the larger defocus blur in the raw image (Fig.~\ref{fig1}b). Scale bars: $50 \mathrm{\mu m}.$}
        \label{fig3_experimental}
\end{figure}

Our experimental setup is schematically depicted in Fig.~\ref{fig1}a, and the experimental results are summarized in Fig.~\ref{fig3_experimental}.
In all experiments, a $532\mathrm{nm}$ CW laser (Oxxius LCX-532S) illuminates a pair of scanning mirrors (Thorlabs GVS012), which are imaged on the back focal plane of an objective lens (OL, $20\times$, $0.4$ NA, Olympus) using a 4-f system. 
The galvanometric mirrors raster-scan the focused beam on the proximal facet of the MCF bundle (Schott P/N 1563385), having an effective imaging area of $0.45 mm$ diameter, containing $\sim 5,600$ cores, with an average core diameter of $4.1 \mathrm{\mu m}$ and an average core-to-core distance of $6.5 \mathrm{\mu m}$. 
The fluorescence signal is filtered by a dichroic mirror (DM) and band-pass filters (F, Semrock BrightLine FF01-576/10-25, and Thorlabs FEL0550), and the descanned image is formed on an sCMOS camera (Andor Zyla 4.2). The target object is composed of Nile Red fluorescent beads (Spherotech), averaging 6$\mathrm{\mu m}$ in diameter, dispersed on a cover glass. A motorized actuator (Thorlabs Z825B) controls the distance between the fiber and the sample.

The ISM endoscopic image is formed in an identical fashion to conventional ISM  image formation \cite{muller2010image}. Specifically, each captured descanned frame is rescaled by a factor of $1/2$ (Fig.~\ref{fig1}d), and its center is shifted to the illumination position. The scaled and shifted frames are summed to form the final ISM image.

To demonstrate the spatial resolution and SNR improvements of ISM over confocal microendoscopy, we image the target at different distances from the fiber distal facet to the target. First, we image the sample at a contact distance ($z_{obj}=0 \mathrm{\mu m}$) to establish the ground truth for the experimental measurements (Fig.~\ref{fig3_experimental}a). Next, we position the target at a distance of $60 \mathrm{\mu m}$, where we observe an improvement in spatial resolution for the ISM-endoscopy image over the confocal image (Fig.~\ref{fig3_experimental}b,c), due to the higher sampling frequency (Fig.~\ref{fig2_sampling}a-b, Fig.~\ref{fig1}d). A cross-section of the three images (Fig.~\ref{fig3_experimental}d) shows the improvement in resolution, resolving the beads that compose the target at this cross-section, as validated in a transmission microscope image of the sample (Fig.~\ref{fig3_experimental}e). 
At a larger distance of $150 \mathrm{\mu m}$ (Fig.~\ref{fig3_experimental}f,g), the ISM image displays a significant improvement in SNR over the confocal image due to the higher collection efficiency, leading to a resolution improvement that does not originate from the higher spatial sampling frequency (Fig.~\ref{fig2_sampling}c,d, supplementary Fig.~S$2$). Fig.~\ref{fig3_experimental}i displays the experimental improvement in collection efficiency (black dots) measured by averaging the fiber cores that image sufficiently bright beads as a function of target distance, together with numerically simulated results. The full information on the numerical simulation is given in Supplementary Section 1.


All displayed ISM, and confocal images are Fourier-interpolated using a Hann filter to eliminate the cores pixelization. Since the ISM image has a $\times 2$ denser grid of 'virtual' cores (Fig.~\ref{fig1}d), the bandwidth of the Hann filter, set by the inverse of the core-to-core pitch, is doubled in the ISM images, compared to the filter used in the confocal image.

\section{Discussion}

In this work, we demonstrated a lensless micro-endoscopy approach that adapts the principles of image-scanning microscopy (ISM) to fiber-bundle endomicroscopy. Our proposed configuration utilizes off-the-shelf components and is based on a conventional confocal fiber-bundle endoscope with a camera replacing the confocal detector. By implementing ISM-endoscopy, we demonstrate an improved collection efficiency of up to a factor of approximately $18$ at an object distance of $150 \mathrm{\mu m}$, and an enhanced spatial resolution compared to confocal endoscopy. 

While the improvement of spatial resolution and collection efficiency in ISM has been extensively studied \cite{muller2010image,sheppard2020pixel}, we highlight an unreported additional advantage of ISM for spatially undersampled systems such as MCF bundles: doubling of the effective spatial sampling frequency, which doubles the accessible spatial frequency bandwidth of the imaging system. 

ISM endoscopy can be applied to any fiber bundle endoscope, and the collection and resolution enhancements may vary according to the MCF parameters and object distances.
This is primarily due to different fiber bundles having different numerical apertures (NA), which affect the distances in which the resolution improvement due to doubling the sampling frequency and improvement in collection efficiency are obtained. 
Importantly, the optimal number of reassigned neighboring cores depends on the object distance and fiber NA, where a larger number of cores collect a useful signal at a larger object distance, and higher NA \cite{sommer2021pixel}.
Combining ISM MCF-based endoscopy with advanced techniques that rely on wavefront shaping for focus-scanning and wavefront correction of fiber distortion \cite{weiss2018two, singh2023multiview} may be interesting for future studies.



A drawback of employing a camera rather than a single detector in the imaging system is a potentially slower image acquisition speed. At larger object distances, it may be interesting to study sparse and/or compressive measurements. 

Lastly, we note that our ISM-based endoscope may enable label-free reflection endomicroscopy, which is challenging to perform in confocal detection. This is because, in reflection endomicroscopy, the illumination and detection wavelengths are identical. Detecting the weak reflected target signals in the presence of relatively-strong background reflections from the MCF facets is a great challenge \cite{choi2022flexible,kang2023fourier}. However, applying ISM while digitally masking the illumination core allows the reconstruction of target images from the signals collected in the neighboring unilluminated cores.

\begin{backmatter}
\bmsection{Funding}
\noindent H2020 European Research Council (101002406), Israel Science Foundation (1361/18).

\bmsection{Disclosures}

\noindent The authors declare no conflicts of interest.

\bmsection{Data availability} 

\noindent Data underlying the results presented in this paper are not publicly available at this time but may be obtained from the authors upon reasonable request.

\end{backmatter}

\bibliography{main}

\newpage

\title{SUPPLEMENTARY MATERIAL}
\author{} 

\setcounter{section}{0}                       

\section{Numerical Simulation parameters and spatial-resolution numerical experiment}
\setcounter{figure}{0}                       
\renewcommand\thefigure{S\arabic{figure}}   

This section describes the parameters used in the simulations of the fiber-scanning techniques presented in Figs.~2-3 in the main text. 
The simulations provide numerically reconstructed images of confocal microendoscopy and ISM-based microendoscopy, together with unsampled (ideal) confocal images. The simulated images are then used to compare the different techniques' spatial resolution and collection efficiency.

The simulation is performed using the experimental parameters of the fiber bundle used in our experiments (Schott P/N 1563385). 
The simulated fiber bundle average distance between adjacent cores is $6.5 \mu m$, with an average core diameter of $4.1 \mu m$ \cite{shekel2020using}. 
The illumination and detection of each fiber core are modeled using an embedded Gaussian beam with a beam quality factor $M^2 = 3$ \cite{siegman1998maybe} and an embedded numerical aperture of $NA_{embedded} = 0.125$, which were chosen to fit the core contact measurements (as in Fig.1b) and the measured collection improvement (Fig.3i). The embedded Gaussian beam has  an effective NA of $NA=M\cdot NA_{embedded} \approx 0.22$. The illumination and detection beams are created using different wavelengths that match the experimental illumination wavelength of $532 \mathrm{nm}$ and the fluorescence emission wavelength of $560 \mathrm{nm}$. 
In the experiment, efforts were made to illuminate mainly the fundamental mode of every fiber core, thus reducing the effective NA of the illumination beam compared to the full NA of the core. Indeed, the embedded Gaussian model NA is lower than the NA of the fiber, as reported in previous studies of $\sim 0.31-0.35$ \cite{weiss2018two,shekel2020using}
The Numerical Aperture is defined as $NA = \lambda / \pi \omega_0$, where $\lambda$ represents the wavelength, and $\omega_0$ denotes the waist radius. 

The simulation is performed as follows: in each illumination step, a single core of the fiber bundle is illuminated in the fundamental mode. The fiber fundamental mode is modeled as an embedded Gaussian beam, $E_{i}(r)$, assuming that the mode can be approximated by a Gaussian beam with a waist that is $M$ times larger and a beam divergence that is $M$ times larger but with the same Rayleigh distance and curvature of a Gaussian beam \cite{siegman1998maybe}. The resulting illumination beam intensity is given from free-space propagation of the fiber mode $P_{i}(r) = \mathcal{P}_{z}{[E_{i}(r)]}$, Where $\mathcal{P}_{z}$ is the free-space angular-spectrum propagation over a distance z from the fiber facet to the target.
Assuming a thin planar target object, the fluorescence signal is calculated as the pointwise product of the illumination beam intensity and the object intensity mask: $I_{fluorescence} = |O(r)|^2 \cdot |P_{i}(r)|^2$, where $O(r)$ denotes the object reflectivity.

\begin{figure}[ht!]
	\centering
	\includegraphics   [width=\textwidth,]
	{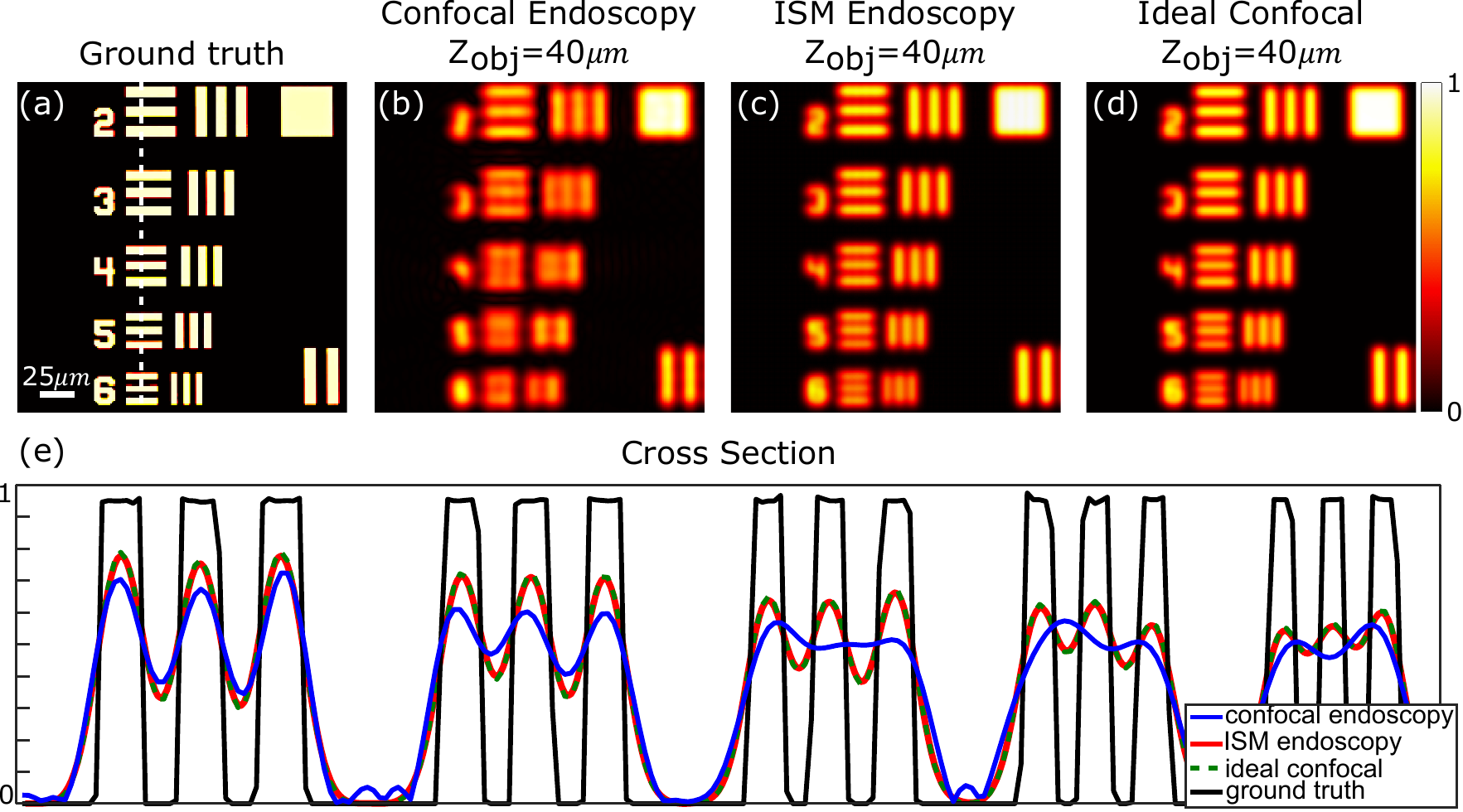}
	\caption{\textbf{Numerical noise-free investigation of the imaging resolution.} (a) the target object is a USAF resolution test target placed at a distance of $40 \mu m$ from the distal facet. (b) confocal bundle-based endoscopy image, (c) ISM bundle-based microendoscopy image, (d)  ideal (sampling-free) confocal microscopy image. (e) cross-sections along the dotted white line in (a) display the improvement in ISM-endoscopy (dotted red line) over confocal-endoscopy (blue line) where elements 4, 5, and 6 are clearly resolved in the final ISM image
 Scale bars: $25 \mu m.$ }
	\label{fig_S1}
\end{figure}

The fluorescence signal reaching the distal facet of the fiber is obtained by convolving the illuminated area of the sample with the point-spread function (PSF): $I_{distal facet} = I_{fluorescence} \ast |P_{i}(r)|^2$. By utilizing reciprocity, this PSF is calculated by propagating the fiber mode (now at the detection wavelength) from the distal facet to the target plane. To simulate fiber sampling, the resulting convolved intensity signal on the fiber facet is convolved with the mode intensity profile of a single core at the fiber facet $|E_{i}(r)|^2$, and is sampled by a grid of the cores-centers positions.  
In the case of confocal endoscopy, only the signal collected by the illumination core is utilized. For ISM endoscopy, all detected signals are collected and allocated to the midpoint between the illumination and detection core. Finally, the intensity images created by illuminating different cores are summed to generate the final ISM and confocal images.
All displayed ISM, and confocal images are Fourier-interpolated to eliminate the cores pixelization. Since the ISM image has a $\times 2$ denser grid of 'virtual' cores (Fig.~1d in the main text), the bandwidth of the filter, set by the inverse of the core-to-core pitch, is doubled in the ISM images, compared to the filter used in the confocal image.
The ideal confocal image, displayed in Fig.~2a and supplementary Fig.~\ref{fig_S1}, is calculated directly by convolving the target with the pointwise product of the illumination and detection PSFs, as given by Eq.~1 in the main text.

Figure~\ref{fig_S1} presents a noiseless numerical comparison of reflective fluorescence USAF target imaging using ISM-endoscopy, confocal endoscopy, and an ideal confocal microscope. The target is $40 \mu m$ away from the fiber facet. The final image obtained by ISM endoscopy (Fig.\ref{fig_S1}b) shows superior resolution than the confocal endoscopy image (Fig.\ref{fig_S1}c) and is similar to the ideal confocal microscope image (Fig.\ref{fig_S1}d). Examining a cross-section over the horizontal lines of the USAF target (Fig.\ref{fig_S1}e, indicated by the dashed white line in Fig.\ref{fig_S1}a), we observe that the ISM image can clearly distinguish between adjacent lines, whereas the confocal endoscopy image cannot.

\newpage

\section{Numerical investigation of resolution improvement in the presence of noise}

\begin{figure}[ht!]
	\centering
	\includegraphics   [width=\textwidth,]
	{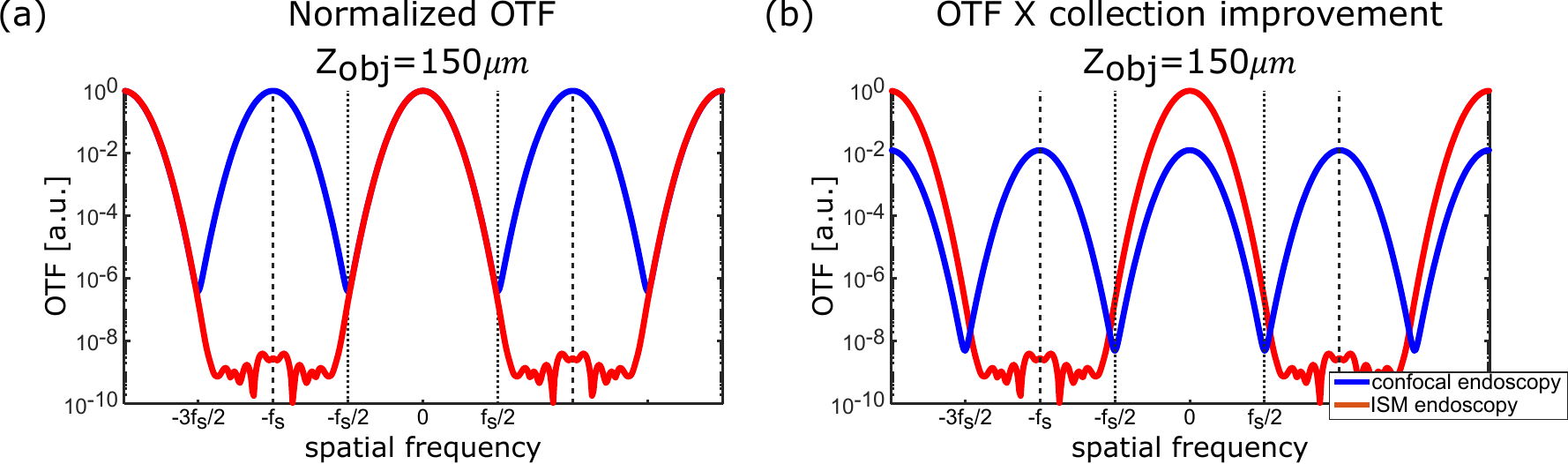}
	\caption{\textbf{Numerical investigation of  OTFs taking into account the collection efficiency improvement.} (a) The optical transfer function (OTF) of ISM-endoscopy and confocal-endoscopy exhibit a similar bandwidth when the collection efficiency is not taken into account. (b) OTFs multiplied by the collection efficiency improvement reveal the resolution improvement possible due to the increase in SNR.}
	\label{fig_S2}
\end{figure}

As mentioned in the main text, the apparent resolution improvement at large target distances (Fig.~3) is not the result of the finer spatial sampling grid, but of higher collection efficiency. This is not reflected in the normalized OTF images presented in Fig.2.
To demonstrate the potential resolution improvement at large distances due to the improvement in collection efficiency, we present in Fig.\ref{fig_S2} the numerically simulated OTFs obtained in confocal and ISM without normalizing by the maximal intensity. Although the optical transfer function (OTF) of ISM-endoscopy and confocal-endoscopy exhibit similar bandwidth in the absence of noise (Fig.~\ref{fig_S2}a), the scaled OTFs (multiplied by the detection improvement, Fig.~\ref{fig_S2}b), show that higher spatial frequency content can be realized in the presence of shot-noise or detector noise.

\end{document}